\documentstyle[12pt]{article}

\begin{document}
\hbadness=10000
\hbadness=10000
\begin{titlepage}
\nopagebreak
\begin{flushright}
{\normalsize
HIP-1999-11/TH\\
March, 1999}\\
\end{flushright}
\vspace{0.7cm}

\begin{center}
{\Large\bf On Different Criteria for Confinement}

\vspace{1cm}

{\bf Masud Chaichian    
} 
and {\bf Tatsuo Kobayashi 
}\\

\vspace{0.5cm}
Department of Physics, High Energy Physics Division\\
    University of Helsinki \\
and\\
Helsinki Institute of Physics\\
    P.O. Box 9 (Siltavuorenpenger 20 C), 
    FIN-00014 Helsinki, Finland \\

(March 11, 1999)\\

\end{center}
\vspace{0.5cm}

\nopagebreak

\begin{abstract}
We compare two approaches in supersymmetric confinement, 
the Seiberg formulation and the superconvergence rule.
For the latter, the critical point is $\gamma_{00}=0$ in the Landau gauge.
We find $4\gamma_{00}=\beta_0$ is the critical point for most of 
confining theories without a tree-level superpotential 
in the Seiberg formulation, in particular, in the large $N_c$ and 
$N_f$ limit.
We show how confining theories with a discrete symmetry and 
a tree-level superpotential connect these two critical points: 
the large and small discrete symmetry limits correspond to 
the critical points $4\gamma_{00}=\beta_0$ and $\gamma_{00}=0$, 
respectively.
\end{abstract}
\vfill
\end{titlepage}
\pagestyle{plain}
\newpage
\def\thefootnote{\fnsymbol{footnote}}

\section{Introduction}

There are many approaches to probe confinement, 
the Nishijima and Oehme-Zimmermann (NOZ) approach, lattice calculations, 
the Schwinger-Dyson equations, the recent Seiberg formulation, etc. 
However, relations between different approaches are not clear.
Indeed, the definition and implication of confinement  
in these approaches differ from each other.
It is important to compare different approaches 
in order to understand possible relations among them.
To this end, we compare two approaches, the Seiberg and 
the NOZ approaches.

In the NOZ approach, the superconvergence rule for the gluon propagator 
has been investigated \cite{conv,conv1,rev1} and 
the approach is related with the Kugo-Ojima confinement 
mechanism \cite{ko}.
The superconvergence for the gluon propagator is realized if 
$Z_3^{-1}=0$,
where $Z_3$ is the renormalization constant of the 
gluon wave function.
In this sense confinement, as well as other nontrivial aspects, implies 
the behavior that some of the renormalization constants, $Z_i$ or 
$Z_i^{-1}$ should vanish.
In this approach the renormalization group (RG) equations play 
an important role.
Its discussion is applicable in the weak coupling region
and one-loop RG equations are reliable.
In the Landau gauge $\alpha =0$, we have $Z_3^{-1}=0$, which implies
the superconvergence, only if 
$\beta_0 < 0$  and $\gamma_{00} < 0$,
where $\beta_0$ and $\gamma_{00}$ are the one-loop $\beta$-function 
coefficient of the gauge coupling $g$ and the one-loop anomalous 
dimension coefficient of the gluon field, respectively.
The NOZ approach is model-independent.
The condition for the superconvergence is applicable 
for gauge theories with any  gauge groups and any matter fields.
Also it is applicable for both non-supersymmetric and 
supersymmetric theories  \cite{oehme}.
Only the $\beta$-function and the anomalous dimension are 
the decisive quantities.

On the other hand, in the Seiberg formulation, holomorphy and 
global symmetries including 
$R$-symmetries play an essential role \cite{seiberg,rev2}.
By use of the power of symmetries, one can discuss nonperturbative 
aspects of supersymmetric gauge theories in the strong coupling region.
In this formulation confinement implies the description 
of the infrared effective theory in terms of gauge invariant operators 
on the moduli spaces.
The formulation is theory-dependent, i.e. the analyses depend on 
global symmetries of theories.
Several types of confining theories without a tree-level 
superpotential \cite{rev3}, i.e. 
the theories with non-vanishing superpotential 
(s-confinement) \cite{csaki}, 
the theories with vanishing superpotential and quantum deformed 
moduli space, the theories with vanishing superpotential and 
an affine moduli space \cite{dotti}, have been found.
Furthermore, the confining theories with 
quantum deformed moduli space are classified into two types: 
in one type the quantum deformed constraint is invariant under the 
global symmetry (i-confinement) and in the other the constraint 
is covariant (c-confinement) \cite{gn}.
In addition, recently several s-confining theories with 
a tree-level superpotential have also been 
found \cite{tree1,tree2,tree3}.

In order to compare the two approaches mentioned above, 
it is useful to 
write a generic confinement condition in the Seiberg 
formulation in terms of $\gamma_{00}$ and $\beta_0$ and 
study its implications 
in RG equations.

Recently, Cs\'aki {\it et al.} have obtained a generic  
viewpoint in the Seiberg approach \cite{csaki}.
They have derived necessary conditions for the s-, i- and 
c-confining theories, 
which are written in terms of Dynkin indices of matter fields.
We investigate the confinement condition 
following Ref.\cite{csaki} and describe it in terms of 
$\gamma_{00}$ and $\beta_0$.
Similarly, we investigate confining theories with a tree-level 
superpotential.
Then, we compare the superconvergence rule and the implications 
of Seiberg's confinement and present 
a conjecture which would fill the gap between them.


\section{Superconvergence rule}

First we review briefly the superconvergence rule.
We begin with the renormalization constants for the gluon field,
$A_\mu^{(0)}=Z_3^{1/2}A_\mu$,  
$\Gamma_{3A}^{(0)}=Z_1\Gamma_{3A}$ and  
$g^{(0)}=Z_g g$, 
where $A_\mu$ is the gluon field and $\Gamma_{3A}$ is the three-gluon 
vertex.
Here the $A_\mu^{(0)}$ denotes the unrenormalized gluon field and 
the superscript $(0)$ in $\Gamma_{3A}^{(0)}$ and $g^{(0)}$ 
means the same.
Obviously, we have the relation $Z_g=Z_1Z_3^{-3/2}$.
The RG equations for these renormalization constants are written as 
${d Z_3^{-1} / d \rho} = -2\gamma Z_3^{-1}$ and 
${d Z_g^{-1} / d \rho} = (\beta/g) Z_g^{-1}, $
where $\beta$ and $\gamma$ are the $\beta$-function of $g$ and 
the anomalous dimension of the gluon field.
Here we restrict ourselves to the Landau gauge $\alpha =0$ \cite{fnote}.
The $\beta$-function and the anomalous dimension are expanded as, 
\begin{eqnarray}
\beta (g) &=& g^3(\beta_0+\beta_1g^2 +\cdots), \\
\gamma (g) &=& g^2(\gamma_{00}+\gamma_{10}g^2 +\cdots).
\end{eqnarray}

In the Landau gauge, the gluon propagator is given by 
\begin{eqnarray}
D_{\mu \nu}(k) &= &\left( g_{\mu \nu} -
{k_\mu k_\nu \over k^2-i\varepsilon } \right) D(k^2), \\
D(k^2) &= & \int dm^2 {\rho(m^2) \over k^2-m^2-i\varepsilon },
\end{eqnarray}
where $\rho(m^2)$ is the spectral function.
$D(k^2)$ follows the same RG equation as $Z_3^{-1}$ and 
we have 
$Z_3^{-1} = \int dm^2 \rho(m^2)$.
Therefore, if $Z_3^{-1}=0$, then we 
have the superconvergence sum rule,
$\int dm^2 \rho(m^2) =0$.

Now let us calculate $Z_3^{-1}$ and $\int dm^2 \rho(m^2)$.
We are interested in the asymptotic free theory, i.e. $\beta < 0$, 
and we consider the weak coupling region where  
the one-loop RG equations are reliable.
In this case, we use the RG equations to calculate $Z_3^{-1}$ as follows,
\begin{eqnarray}
Z_3^{-1} 
&=& \exp \left( \int^\infty_0d \rho 2\gamma \right),
\nonumber  \\
&=& \exp \left( \int^{g_{\infty}}_{g(0)}dg\beta_g^{-1} 2\gamma
\right), \nonumber \\
&=& (g_{\infty}/g(0))^{2\gamma_{00}/\beta_0},
\label{Z3}
\end{eqnarray}
where $g(0)$ is a nonvanishing finite value at $\rho=0$.
Because of the asymptotic freedom, we have the behavior
$g_{\infty} \rightarrow 0$
in the limit $\rho \rightarrow \infty$.
Hence, we have $Z_3^{-1}=\int dm^2 \rho(m^2)=0$ 
if $\gamma_{00} <0$.
Thus, the critical point for the superconvergence is 
\begin{equation}
\gamma_{00} =0.
\label{gamma00}
\end{equation}
In the NOZ approach it is important to investigate the 
behavior of the renormalization constants, i.e. to calculate 
$Z_i$ or $Z_i^{-1}$ and investigate critical points of these values.
On the other hand, Kugo conjectured that $Z_1Z_3^{-1}=0$ is a sufficient 
condition for the color confinement \cite{kugo}.

Note that the NOZ approach is generic and the superconvergence 
rule is applicable for both 
supersymmetric and non-supersymmetric gauge theories with any gauge group 
and matter fields in any representations.
Furthermore, the approach focuses on the weak coupling region 
implying that one could derive nontrivial results 
even from knowledge at the weak coupling region.

\section{The Seiberg confinement}

In the Seiberg formulation, the probe of confining theories implies 
investigation 
of the superpotential of gauge invariants together with 
their moduli space and constraints.
Several types of confining theories have been found \cite{rev3}, 
s-confinement \cite{csaki}, 
i-confinement, c-confinement \cite{gn} and affine 
confinement \cite{dotti}.

For example, in the supersymmetric QCD with 
the gauge group $SU(N_c)$ and $N_f$ flavors of quark pairs, 
the case with $N_f=N_c$ corresponds to 
the i-confinement, while the $N_f=N_c+1$ case corresponds to the 
s-confinement.
In the supersymmetric QCD, the flavor number $N_f=N_c+1$ is a critical 
point. 
For a larger flavor $N_f > N_c+1$ we do not have confinement.

In Ref.\cite{csaki} necessary conditions for the s-, i- and 
c-confinements have been derived without the presence of a tree-level 
superpotential.
Let us consider the supersymmetric gauge theory with the gauge group 
$G$ and $N$ chiral matter multiplets $\Phi_i$ ($i=1,\cdots, N$).
We define $N$ $R$-symmetries $U(1)_{Ri}$ ($i=1,\cdots, N$) such that 
for the $i$-th $R$-symmetry $U(1)_{Ri}$ all chiral matter fields 
except the $i$-th chiral multiplet $\Phi_i$ have vanishing $R$-charges 
and $\Phi_i$ has the $R$-charge $q_i$.
These $R$-symmetries should have vanishing gauge anomaly.
Thus, we fix the $R$-charge $q_i=\Delta/\mu_i$, where 
\begin{equation}
\Delta \equiv \sum_{j=1}^{N}\mu_j - \mu_G,  
\end{equation}
and $\mu_i$ is the Dynkin index for the gauge representation of 
$\Phi_i$ and $\mu_G$ is the index of the adjoint representation.
Here we follow the normalization used in Ref.\cite{csaki}, where 
for instance the fundamental representations of the $SU(N)$ group have 
the Dynkin index $\mu_{fund}=1$.

The superpotential should have the $R$-charge 2 for $U(1)_{Ri}$.
Hence, the superpotential must be a combination of terms of the form
$\Lambda^3 \prod 
(\Phi_i/\Lambda)^{2\mu_i/\Delta}$,
where $\Lambda$ is the dynamical scale.
For the s-confinement we must have the smooth superpotential, i.e.  
the exponents of $\Phi_i$ ($i=1,\cdots, N$) should be positive integers.
This implies $\Delta =2$ or 1.
Only the former one is available in the present 
normalization \cite{csaki}, i.e. 
\begin{equation}
\Delta =2 {\rm ~~~for~~~s-confinement}.
\label{scon}
\end{equation}
Similarly, we can discuss the i- or c-confinement, where 
we have the vanishing superpotential.
We have the condition,
\begin{equation}
\Delta =0 {\rm ~~~for~~~i-~and~c-confinement}.
\label{iccon}
\end{equation}
Eqs. (\ref{scon}) and (\ref{iccon}) are the necessary conditions for 
the s-confinement and i- and c-confinement, respectively.
Indeed, in the supersymmetric QCD we find $\Delta =2$ for 
$N_f=N_c+1$ and $\Delta =0$ for $N_f=N_c$.


Here we study implications of the conditions 
(\ref{scon}) and (\ref{iccon}) from the viewpoint of the NOZ approach.
We consider the weak coupling region where the one-loop RG equations 
are reliable.
We describe the conditions (\ref{scon}) 
and (\ref{iccon}) in the plane $(\beta_0,\gamma_{00})$.
In supersymmetric gauge theory, $\beta_0$ and 
$\gamma_{00}$ are obtained as 
\begin{eqnarray}
(16\pi^2)2\beta_0 &=& -3\mu_G+\sum_{i=1}^N\mu_i = -2 \mu_G+\Delta, 
\label{rg1}\\
(16\pi^2)2\gamma_{00} &=& -{3 \over 2}\mu_G+\sum_{i=1}^N\mu_i
= -{1 \over 2} \mu_G+\Delta.
\label{rg2}
\end{eqnarray}
Here we use eq. (\ref{iccon}), so that 
we find the relation between $\beta_0$ and $\gamma_{00}$, 
\begin{equation}
4\gamma_{00} = \beta_0,
\label{4gb}
\end{equation}
for the i- and c-confinement.
We have 
$4\gamma_{00}=\beta_0+3/(16\pi^2)$ for the s-confinement.
In the large $\mu_G$ limit, e.g. the large $N_c$ limit of the gauge group 
$SU(N_c)$, it becomes eq.(\ref{4gb}).
The large $\mu_G$ limit also implies the large flavor number limit through 
eqs.(\ref{scon}) and (\ref{iccon}).

In the affine confining theories \cite{gn}, 
the universal index constraint is no longer applicable.
However, in the large $\mu_G$ limit the relation (\ref{4gb}) is 
realized for the theories explicitly obtained 
in Ref.\cite{gn} \cite{fnote1}.
Furthermore, the other type of confining theory has been found, 
i.e. the $SO(N)$ gauge theory with $(N-3)$ vectors \cite{is}.
This theory is not classified into any type of the above.
This theory has $2\sum_j\mu_j < \mu_G$ and $2\sum_j\mu_j = \mu_G$ 
in the large $N$ limit.
Hence, in such limit we have 
$5/2 \gamma_{00} = \beta_{0}$.

As a result eq. (\ref{4gb}) in the $(\beta_0,\gamma_{00})$ plane 
is significant 
for most of the Seiberg type of confining theories, in particular 
i- and c- confinement and s- and affine confinement for large $\mu_G$.
It is different from the critical point of the superconvergence 
(\ref{gamma00}).
By performing calculations similar to eq.(\ref{Z3}), we find that 
eq.(\ref{4gb}) is the critical point for the value $Z_1^{-1}Z_3^{-1/2}$, 
i.e. $Z_1^{-1}Z_3^{-1/2}=0$ if $4\gamma_{00} < \beta_0$.
Similarly, $Z_1^{-1}$ vanishes if $3\gamma_{00} < \beta_0$.
These regions require less matter fields than $\gamma_{00} <0$.
Therefore, the single condition $Z_3^{-1}=0$ appears not to be sufficient 
for the Seiberg type of supersymmetric confining theories.

Before we consider the gap between eqs.(\ref{4gb}) and (\ref{gamma00}) 
through investigation of a tree-level superpotential, 
we give a comment on a recent lattice calculation \cite{iwasaki}, which 
shows that there is a confinement phase for 
the non-supersymmetric QCD with the gauge group $SU(3)$ and 
6 or less flavors of quark pairs.
Amusingly, this coincides with the region with $3\gamma_{00} < \beta_0$ 
for the non-supersymmetric case.


Up to now, we have considered confining theories without a tree-level 
superpotential.
In particular, eq.(\ref{4gb}) is significant for the i- and c-
confinement and s- and affine confinement for large $\mu_G$.
Now we discuss confining theories with a tree-level superpotential.
Such a consideration is suggestive for understanding the origin of 
the gap between 
eq.(\ref{4gb}) and eq.(\ref{gamma00}) as we shall see.

Recently, several s-confining theories with a tree-level superpotential 
have been discussed \cite{tree1,tree2,tree3}.
These new s-confining theories are shown in Table 1.
They have $N_f$ flavors of fundamental pairs for the 
gauge groups $SU(N_c)$ and $Sp(2N_c)$ and $N_f$ flavors of 
vector representations for the gauge group $SO(N_c)$.
In addition, they contain additional tensors and the tensors have 
a tree-level superpotential.
The presence of a tree-level superpotential breaks explicitly some of 
$U(1)$ symmetries which play an essential role in the index constraint.
Thus, neither eq.(\ref{scon}) nor (\ref{iccon}) holds.
Instead, these theories possess discrete symmetries which are shown in 
Table 1.

For example, let us take the second theory in Table 1, which has  
the gauge group $SU(N_c)$, $N_f$ flavors of quark pairs, an additional  
antisymmetric tensor multiplet and its conjugate $X$ and $\bar X$, and 
the tree-level superpotential $W_{\rm tree}={\rm Tr} (X\bar X)^{k+1}$ 
\cite{tree2}.
This theory has the discrete symmetry $Z_{2(k+1)N_f}$.
If $N_c=(2k+1)N_f-4k-1$, s-confinement is realized.
The theory with one flavor less has a quantum modified moduli space.
In the second s-confining theory we have 
$\Delta =2(N_f-2)$ and  $\mu_G = 2(2k+1)N_f-8k-2$ 
as shown in Table 1.
As the discrete symmetry becomes large, i.e. $k$ increases, 
the relation between $\gamma_{00}$ and $\beta_0$ becomes close to 
eq.(\ref{4gb}).
On the other hand, as the discrete symmetry becomes small, 
the ratio $\beta_0 /\gamma_{00}$ increases.
To simplify the discussion, here we take the large $N_f$ limit 
as we did previously. 
We denote the discrete symmetry as $Z_{(m+1)N_f}$.
In this case, we have 
\begin{equation}
\Delta =2N_f, \qquad \mu_G = 2mN_f.
\end{equation}
Indeed these relations in the large $N_f$ limit always hold in any 
s-confining theory with the discrete symmetry $Z_{(m+1)(N_f+\ell)}$ 
among the theories shown in Table 1.
Thus, we can obtain a common relation, 
\begin{equation}
{\beta_0 \over \gamma_{00}} = {-4m +2 \over -m +2},
\label{comm}
\end{equation}
for the large $N_f$ limit.
As the discrete symmetry and $m$ become small, 
the ratio $\beta_0/\gamma_{00}$ becomes large, i.e. $\gamma_{00}$ 
approaches the point  $\gamma_{00}=0$ 
from the negative side for $\beta_0$ fixed.
Actually for $m=2$ we have $\gamma_{00}=0$, which is the critical point 
for the superconvergence rule (\ref{gamma00}).
Note that the discrete symmetry corresponding to $m=1$ is realized 
only for the first, sixth and eighth theories in Table 1, but in this case 
the superpotential corresponds to a mass term of $X$.
After integrating out the heavy mode $X$, we are left with 
no superpotential.
Thus, $m=2$ is the minimum value.
On the other hand, as the discrete symmetry and $m$ become large, 
the ratio approaches the value $\beta_0/\gamma_{00}=4$.
The relation (\ref{4gb}) is realized at the limit 
$Z_{(m+1)N_f} \rightarrow Z_{\infty}$.
This aspect is quite suggestive and seems to be of significance.
We have shown that there is indeed a gap for 
the critical values of $(\gamma_{00},\beta_0)$ between the 
superconvergence rule and the Seiberg type of confining theories without 
a tree-level superpotential.
The former critical point is $\gamma_{00}=0$, while the latter one 
is $4\gamma_{00}=\beta_0$.
However, the above theories with a superpotential and 
the discrete symmetry $Z_{(m+1)N_f}$ provide with a bridge over the gap.
The small and the large $Z_{(m+1)N_f}$ limits correspond to 
$\gamma_{00}=0$ and $4\gamma_{00}=\beta_0$, respectively.
That would imply that below $\gamma_{00}=0$ confinement could, 
in principle, happen, but large global symmetries prevent 
its occurrence.
Thus, in order to find out what else, besides the superconvergence rule, 
is needed, specific types of global symmetries should be further 
investigated.
This would provide with a physical understanding of the relation 
between the different approaches to confinement.

\section{Conclusion}

We have studied the supersymmetric confining theories already obtained 
in the Seiberg formulation from the NOZ viewpoint.
We have described the confinement conditions 
in terms of $\beta_0$ and $\gamma_{00}$.
The region around $4\gamma_{00}=\beta_0$ seems to be significant.
The single condition $\gamma_{00} <0$, i.e. $Z_3^{-1}=0$ 
is not sufficient for the Seiberg type of supersymmetric confining 
theories.
There is a gap between the critical points for the Seiberg 
type of supersymmetric confining theories and the superconvergence 
rule, i.e. $4\gamma_{00}=\beta_0$ and $\gamma_{00}=0$.

We have also considered confining theories with the discrete 
symmetry $Z_{(m+1)N_f}$ and a tree-level superpotential.
In this type of theories, the ratio $\beta_0/\gamma_{00}$ is close to 4 
for large $m$ and the ratio increases as $m$ decreases.
Thus, the large discrete symmetry $Z_{(m+1)N_f}$ corresponds to the 
critical point $4\gamma_{00}=\beta_0$ and the small discrete symmetry 
corresponds to the critical point for the superconvergence rule 
$\gamma_{00}=0$.
This strongly suggests that what makes the gap between $4\gamma_{00}=\beta_0$ 
and $\gamma_{00}=0$ is the existence of global symmetry.

Here we have considered the confining theories with the discrete 
symmetry $Z_{(m+1)N_f}$ and a tree-level superpotential, which have 
been previously obtained.
Although we have derived the common relation (\ref{comm}) from 
the consideration of explicit theories of Table 1, we are inclined 
to suggest that this relation could hold in general.
Therefore, it is of importance to extend further the analyses 
to more general types of theories with discrete symmetries.


\section*{Acknowledgments}  
The authors are very grateful to Kazuhiko~Nishijima for 
various enlightening discussions and valuable remarks.
This work was supported by the Academy of Finland under 
Project no. 44129.

\newpage
\hskip -2cm
\begin{tabular}{|c|c|c|c|c|}
\hline
 & 1 & 2 & 3 & 4 \\ \hline
Gauge group & $SU(N_c)$ & $SU(N_c)$ & $SU(N_c)$ & $SU(N_c)$ \\
tensors~$X$ ($\bar X$)   & adj.      & asym. + $\overline{\rm asym.}$  & 
sym. + $\overline{\rm sym.}$ & asym. + $\overline{\rm sym.}$ \\
$W_{\rm tree}$ & $X^{k+1}$ & $(X\bar X)^{k+1}$ & $(X\bar X)^{k+1}$ & 
$(X\bar X)^{2(k+1)}$ \\
discrete & & & & \\
symmetry &  $Z_{(k+1)N_f}$ &  $Z_{2(k+1)N_f}$ &  $Z_{2(k+1)N_f}$ &  
$Z_{4(k+1)(N_f+4)}$ \\
$\Delta$ & $2N_f$ &  $2(N_f-2)$ &  $2(N_f+4)$ & $2(N_f+5)$ \\
$\mu_G/2$ & $(kN_f-1)$ & $(2k+1)N_f-4k-1$ & $(2k+1)N_f+4k-1$ & 
$(4k+3)(N_f+4)-1$ \\ \hline
\end{tabular}

\vskip 1cm
\begin{tabular}{|c|c|c|c|c|}
\hline
 & 5 & 6 & 7 & 8 \\ \hline
Gauge group & $Sp(2N_c)$ & $Sp(2N_c)$ & $SO(N_c)$ & $SO(N_c)$ \\
tensors~$X$    & adj.      & asym.   & adj. & sym.  \\
$W_{\rm tree}$ & $X^{2(k+1)}$ & $X^{k+1}$ & $X^{2(k+1)}$ & $X^{k+1}$ \\
discrete & & & & \\
symmetry &  $Z_{2(k+1)N_f}$ &  $Z_{(k+1)N_f}$ &  $Z_{2(k+1)N_f}$ &  
$Z_{(k+1)N_f}$ \\
$\Delta$ & $2N_f$ &  $2(N_f+2)$ &  $2N_f$ & $2(N_f+6)$ \\
$\mu_G/2$ & $(2k+1)N_f-2$ & $k(N_f-2)$ & $(2k+1)N_f+1$ & 
$k(N_f+4)-3$ \\ \hline
\end{tabular}

\vskip 1cm
Table 1: Confining theories with a tree-level superpotential.
Additional tensors are shown in the third row, 
where ``sym.'' and ``asym.'' 
denote symmetric and antisymmetric representations, respectively, while  
``$\overline{\rm sym.}$''  and ``$\overline{\rm asym.}$'' denote their 
conjugates. 
$N_f$ denotes the flavor number of fundamental multiplets 
and its conjugates 
for $SU(N_c)$ and $Sp(2N_c)$, and vector multiplets for $SO(N_c)$.


\begin{thebibliography}{99}
\bibitem{fnote}
Other gauges have been considered in Ref.\cite{conv3} 
with no significant differences for our further analysis, which 
can be performed in a similar way.
\bibitem{fnote1}
For some of affine confining theories, the confinement is realized only 
for certain gauge groups, but we can not take the large $\mu_G$ limit.
For them, the above statement does not hold.
In particular, one unique theory among them is the $SU(2)$ gauge theory 
with one chiral 
multiplet in the spin 3/2 representation of $SU(2)$ \cite{iss}.
It has the positive $\gamma_{00}$ and thus the condition 
$\gamma_{00} <0$ is not even satisfied.
\end{thebibliography}
\end{document}